\documentclass[12pt,english,preprint]{revtex4-1}
\usepackage[T1]{fontenc}
\usepackage[latin9]{inputenc}
\usepackage{array}
\usepackage{multirow}
\usepackage{amssymb}
\usepackage{graphicx}
\usepackage{subscript}
\usepackage{color}
\usepackage{lineno}

\makeatletter

\providecommand{\tabularnewline}{\\}

\@ifundefined{textcolor}{}
{%
 \definecolor{BLACK}{gray}{0}
 \definecolor{WHITE}{gray}{1}
 \definecolor{RED}{rgb}{1,0,0}
 \definecolor{GREEN}{rgb}{0,1,0}
 \definecolor{BLUE}{rgb}{0,0,1}
 \definecolor{CYAN}{cmyk}{1,0,0,0}
 \definecolor{MAGENTA}{cmyk}{0,1,0,0}
 \definecolor{YELLOW}{cmyk}{0,0,1,0}
}

\makeatother

\usepackage{babel}
\begin{document}
\author{Grigory Kolesov}
\affiliation{John Paulson School of Engineering and Applied Sciences, Harvard University, Cambridge MA}
\author{Boris A. Kolesov}
\affiliation{Nikolaev Institute of Inorganic Chemistry, SB RAS  Novosibirsk, Russia}
\affiliation{Novosibirsk State University, Novosibirsk, Russia}
\author{Efthimios Kaxiras}
\thanks{corresponding author}
\affiliation{Department of Physics, Harvard University, Cambridge, MA}

\title{
Polaron-induced phonon localization and stiffening in
rutile TiO$_{2}$}

\date{\today}
\begin{abstract}
Small polaron formation in transition metal oxides, like the prototypical
material rutile TiO$_2$, remains a puzzle and a challenge to simple theoretical
treatment. In our combined experimental and theoretical study, we examine this
problem using Raman spectroscopy of photo-excited samples and real-time
time-dependent density functional theory (RT-TDDFT), which employs Ehrenfest
dynamics to couple the electronic and ionic subsystems.  We observe
experimentally the unexpected stiffening of the $A_{1g}$ phonon mode under UV
illumination and provide a theoretical explanation for this effect.  Our
analysis also reveals a possible reason for the observed anomalous
temperature-dependence of the Hall mobility.  Small polaron formation in rutile
TiO\textsubscript{2} is a strongly non-adiabatic process and is adequately
described by Ehrenfest dynamics at time scales of polaron formation.

\end{abstract}
\maketitle

\section{Introduction.}

A fundamental unsolved problem in polar materials is how electric charge
carriers are generated and how they move through the solid. In these materials
electrons deform the highly polarizable crystal lattice and form
quasi-particles (polarons) consisting of the electron and a lattice deformation
associated with it. In the case of so-called small polarons the strong and
complicated electron-lattice interaction renders the usual electronic
band-structure description of the charge carriers insufficient.  Polarons have
also proved important in surface electron transfer processes and photocatalysis
on titania where, due to their high binding energy, they act as electron
scavengers~(Refs. \citenum{henderson_surface,kenji2016,kolesov2015anatomy} and
references therein). 

Since the 1960s, the availability of high-quality samples and the absence of
complex magnetic effects have made rutile titania the prototypical
polaron-forming transition metal oxide (TMO).  Still, the basic properties of
the polaron remain controversial. For instance, reported  effective mass values
range from 2 -- 150 $m_{e}$ and room-temperature drift mobility ($\mu_{\perp}$)
values from 0.03 -- 1.4
cm\textsuperscript{2}$\cdot$V\textsuperscript{-1}$\cdot$s\textsuperscript{-1}
~\cite{bogomolov1967,austin_mott,hendryTHz,yan2015,rutile_me1}. The temperature
dependence of mobility also remains subject of a debate: the drift mobility increases
with rising temperature for $T\gtrsim$300 K, while the Hall mobility decreases.
Non-adiabatic polaron theory predicts that both mobilities should rise with
temperature due to thermally-activated hopping mechanisms, while adiabatic
theory predicts the opposite because of higher rate of scattering
events~\cite{austin_mott,holstein1959p1,holstein1959p2,emin,lang_firsov,devreese_review}.
As a possible explanation it was suggested\cite{book_alexandrov1995} that activation
temperature for non-adiabatic hopping is higher in the case of Hall motion, and
increase of the Hall mobility should be expected at higher temperatures.
However earlier experiments in Ref. \citenum{bransky1969hall} where the Hall
mobility was shown to decrease within all measured temperature range up to
$T=1250$~K do not agree with this model.  Here we report yet another puzzling
experimental observation, a phonon-stiffening effect, which is not captured by
existing polaron theories that typically predict softening of the phonon modes
\cite{migdal,rashba1957,alexandrov_strong,alexandrov_intermediate,emin}. 

The formation of small polarons in titania was first
reported by Bogomolov \emph{et al.} \cite{bogomolov1967,bogomolov_optical},
who estimated the lower bound of the polaron binding energy to be
$\sim$0.4 eV. Recent computations support these findings\cite{deskins,rpapolaron}, 
namely that formation of small polarons in rutile titania is 
energetically favored with a binding energy estimated at
$\gtrsim$0.5 eV. The small
polaron view has been challenged~\cite{hendryTHz} while other
computational studies reported smaller polaron binding energies of
$\leq$0.15 eV \cite{setvin,janotti2013dual}.


Here we study the properties of polarons in rutile TiO$_{2}$
both experimentally and computationally. We use Raman spectroscopy
on undoped samples excited with UV laser light. We model the polaron using density
functional theory (DFT) and probe the dynamics of its formation with real-time
time-dependent density functional theory (RT-TDDFT) 
which employs mean-field classical-ion (Ehrenfest) dynamics to couple the electronic
and ionic subsystems\cite{kolesov2015real}. We show that this approach 
is adequate to capture the dynamics and time scales of formation of small polarons
by photo-generated or injected carriers in rutile TiO$_{2}$. In both
experiment and theory we observe stiffening of the \emph{A}\textsubscript{\emph{1g}}
phonon mode in titania upon UV illumination and upon injection of
electrons.  The key elements that lead to this effect are the strong electron
interactions in \emph{d}-shells of the TM atoms and, closely related to it, the
large anharmonic effects of their coupling to the lattice. Based on our results
we propose a qualitative model that explains the temperature dependence of the
Hall mobility.

\section{Methods}
\paragraph*{Experimental.}

For the experimental study, we obtained an undoped monocrystalline
rutile TiO$_{2}$ sample from SurfaceNet GmbH. We measured the Raman
spectra using a LabRAM HR Evolution (Horiba) spectrometer with the
excitation induced by the 632.8 nm line of a He-Ne laser and the 325.03
nm line of a He-Cd laser. The spectra at room temperature were obtained
in the backscattering geometry using a Raman microscope. The diameter
of the laser beam spot on the sample surface was around 1-2 $\mu$m.
The measurements were performed with a spectral resolution of 0.7
cm\textsuperscript{-1} with the 632.8 nm line and 4 cm\textsuperscript{-1}
with the 325.03 nm line.

\paragraph*{Computational.}

For the theoretical modeling, in both the adiabatic and non-adiabatic
(Ehrenfest dynamics) calculations we use our code TDAP-2.0\cite{kolesov2015real},
which is based on the SIESTA package\cite{siesta02,siesta_ldau},
an efficient DFT code employing numerical atomic orbitals (NAO) as
the basis set. Following Refs. \cite{anisimov_polaron,deskins,rpapolaron}
in this work we use the LDA$+U$ approach in its spherically-averaged
form\cite{dudarev}, for computational efficiency. In the non-adiabatic
part we use TD-DFT and mean-field (Ehrenfest) propagation of classical
ions:

\begin{equation}
M_{J}\frac{\partial^{2}\mathbf{R}_{J}}{\partial t^{2}}=-\mathbf{\nabla}_{\mathbf{R}_{J}}
E_{KS}
\end{equation}
where $J$ is the ion index and $E_{KS}$ is the expectation value of electronic
energy. The electronic wavefunctions are propagated with effective
single-particle TD-DFT equations\cite{Runge:1984us}:

\begin{equation}
i\frac{\partial\phi_{n}(t)}{\partial t}=\hat{H}_{KS}[\rho](t)\phi_{n}(t),\label{eq:TDKS}
\end{equation}
with $\phi_{n}$ being the single-particle Kohn-Sham orbitals, $\hat{H}_{KS}$
the Kohn-Sham effective Hamiltonian operator and $\rho\left(\mathbf{r},t\right)=\sum_{n=1}^{N}|\phi_{n}(\mathbf{r},t)|^{2}$
is the electronic density of the system containing $N$ electrons. 

To calculate phonon modes and frequencies we used finite atom displacements
to construct the dynamical matrix (frozen-phonon method). In all cases
described below phonon modes were calculated after complete lattice
relaxation with force tolerance 0.02 eV/\AA.

\emph{Simulation setup. }We used a $4\times4\times4$ rutile titania
supercell in all our calculations. This cell contains 384 atoms and was found
by converging difference between the energies of neutral and charged
($-|q_{e}|$) supercells to within 0.1 eV (see Table S1). Brillouin zone was
sampled at $\Gamma$-point, except for smaller supercells used in convergence
tests where we used Monkhorst-Pack $k$-point grids starting from
$8\times8\times12$ for a single unit cell. As in the works prior to this
\cite{tritsaris2014dynamics,kolesov2015anatomy,kolesov2015real}, we used
PBE$+U$ functional with $U=4.2$ eV in all our calculations.  In this study we
used standard SIESTA pseudopotentials and SZP basis set. In all simulations
that involved an extra electron or electron-hole pair, we employed
spin-polarized version. In Ehrenfest RT-TDDFT simulations we used a time step
$\Delta t=1$ a.u. ($\approx24$ as). We thermalized the system for 1 ps ($T=300$
K) with standard BOMD using 1 fs time steps.

\section{Results}

\begin{figure}
\includegraphics[width=8.6cm]{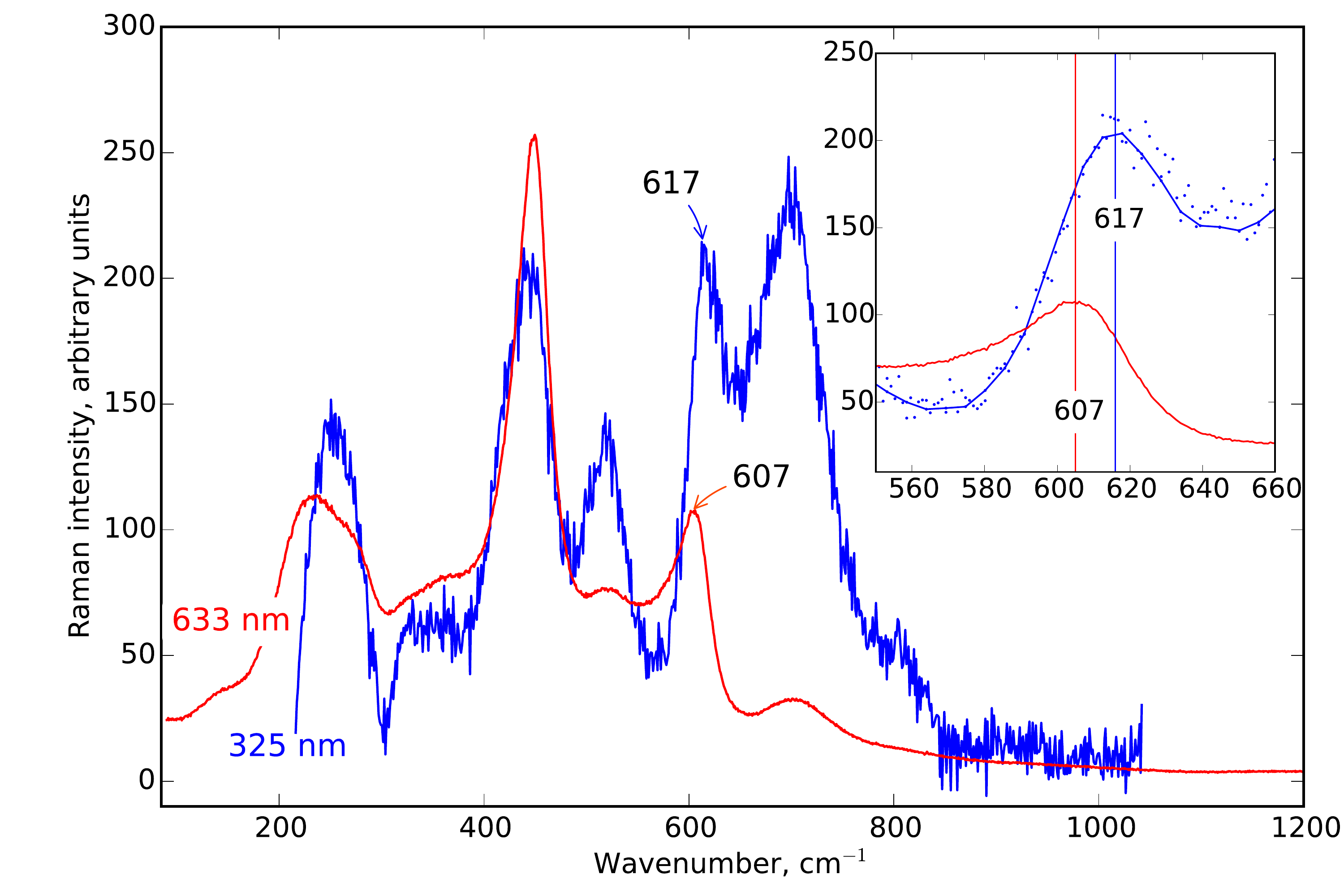}

\protect\caption{Raman spectra of rutile titania
taken with the 632.8 nm laser (red line)  and the 325 nm UV-laser (blue line)
with $\hbar\omega=$3.8 eV $>E_{gap}$. Peak positions in experiment
were assigned by deconvolution in Lorentzian functions (see Fig. S1). The inset
shows a magnified view of the spectra around the 607 cm\protect\textsuperscript{-1}
and 617 cm\protect\textsuperscript{-1}peaks. \label{fig:raman}}
\end{figure}
\begin{figure*}
\includegraphics[width=8.6cm]{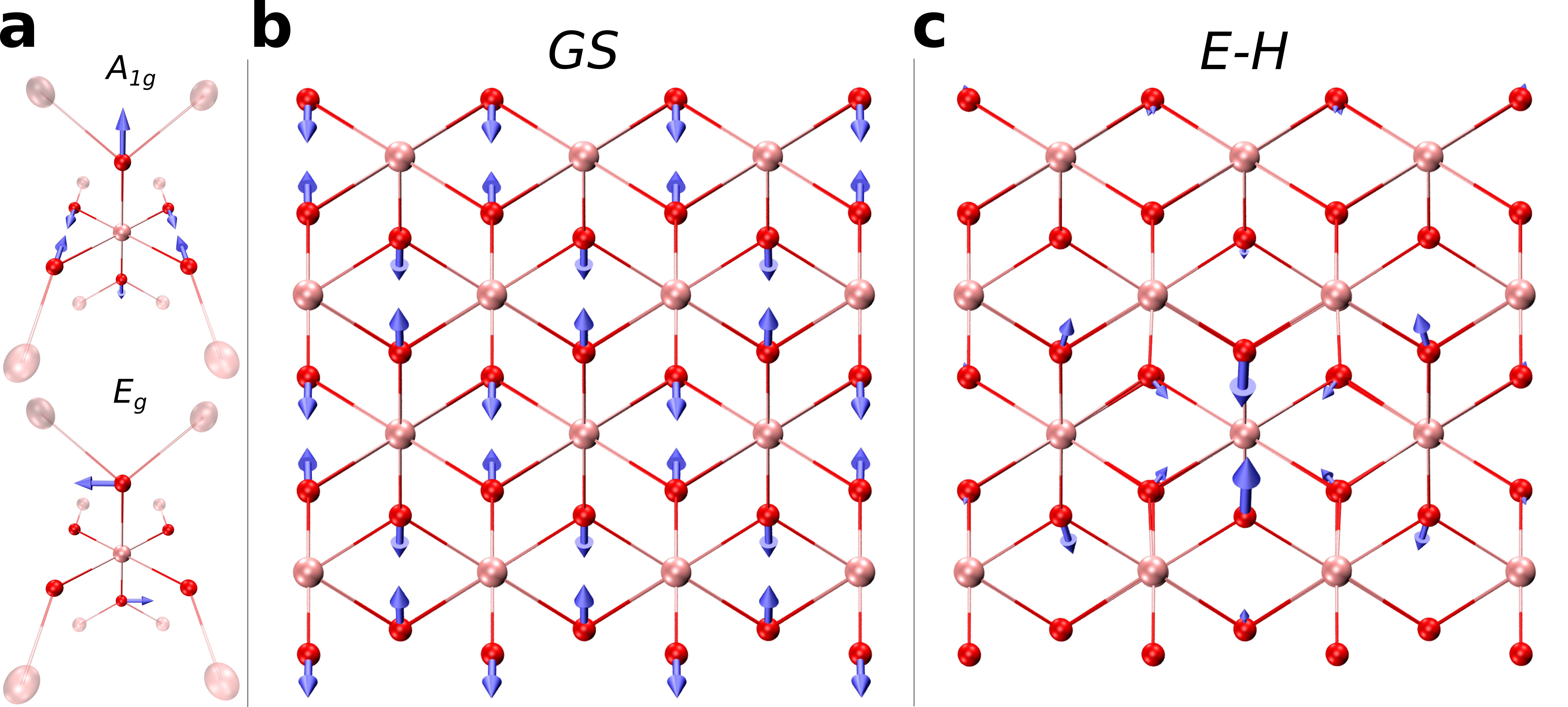}

\protect\caption{(\textbf{a}) 
Stereographic representation of atomic 
displacements corresponding to the $A_{1g}$ and $E_{g}$ phonons
in the ground state (defect-free) bulk titania. 
(\textbf{b})\emph{ } The $A_{1g}$ 607 cm\protect\textsuperscript{-1}
phonon displacement in orthographic projection, in the ground state (\emph{GS} titania.
(\textbf{c}) The localized $A_{1g}^{*}$ phonon at 616 cm\protect\textsuperscript{-1}
in the \emph{E-H} simulations (with the corresponding phonon at
611 cm\protect\textsuperscript{-1}
in the \emph{E}-simulation
being nearly identical, see Fig. S2). \label{fig:locA1g}}
\end{figure*}

Rutile TiO\textsubscript{2} has four Raman-active modes with symmetries
$A_{1g}$, $E_g$, $B_{1g}$ and $B_{2g}$ \cite{porto1967raman}. In Fig.
\ref{fig:raman} we present Raman spectra obtained in the normal state of the
sample (red line) and after it was photo-excited with UV laser (blue line).
The peaks at 440 and 607 cm\textsuperscript{-1} correspond to the $E_{g}$ and
$A_{1g}$ modes respectively\cite{porto1967raman,tio2_raman}, shown schematically in Fig.
\ref{fig:locA1g}a.  The peak at 143
cm\textsuperscript{-1} ($B_{1g}$ phonon) is obscured by the combination peak at
235 cm\textsuperscript{-1} \ \cite{porto1967raman} and can not be seen in the
blue line because it is below the resolution range of the spectrometer used in
the UV experiment. The peak at 823 cm\textsuperscript{-1} which corresponds to
the $B_{2g}$ mode is known to be exceptionally weak and difficult to
resolve\cite{porto1967raman,tio2_raman} and is not seen in the spectra
presented here.  The most interesting feature of the two spectra is a shift of
the $A_{1g}$ 607 cm\textsuperscript{-1} peak to the higher wavenumber of 617
cm\textsuperscript{-1 }in the blue spectrum.  Typically at higher temperatures
of the sample a shift of the peaks to lower wavenumber is
expected\cite{tio2_raman}.  Thus, the shift of the $A_{1g}$ mode to higher
wavenumber cannot be explained by heating of the sample due to carrier
relaxation to the band edges, but could be caused by polaron-induced structural changes.  


Our calculation of the lowest charge-neutral excited state within the DFT
$\Delta$SCF method \cite{kolesov2015real} yielded excitation energy
$E_{g}^{\Delta SCF}=3.7$ eV in the bulk at $T$=0 K. After correcting this
energy for the spin contamination, $E_{g}\approx 2 E_{g}^{\Delta
SCF}-E_{triplet}$ \ \cite{Ziegler1977,kowalczyk_dscf2016}, we obtained
$E_{g}=3.4$ eV. This agrees very well with the value obtained from BSE
calculations (3.3 eV)~\cite{hybertsen}, thus confirming our choice of the value
$U=4.2$ eV, but overestimates the experimental band gap of 3.03 eV. 
%
We calculate the polaron binding energy
$E_{p}$ from the difference of total electronic energies
of the charged system (with excess charge of $-e$) of the 
ideal  bulk $4\times4\times4$ supercell and the relaxed supercell with the same charge. Because
of the high dielectric constant of titania and large supercell size, we disregarded energy
corrections arising from charge interaction between replicas and the
compensating uniform positive background charge that is usually introduced
to obtain converged energies of charged systems in periodic-cell
conditions. 
We find $E_{p}\approx0.9$ eV. 
The \emph{lower} bound for $E_{p}$ in rutile titania was previously estimated from
experimental data in the seminal work by Bogomolov\emph{ et al.}~\cite{bogomolov1967}  
to be $\sim0.4$ eV, and by Austin and Mott~\cite{austin_mott} to be $\sim0.6$
eV, with authors suggesting that this value is likely to be underestimated. Our
value of $0.9$ eV is in reasonable agreement with these estimates, as well as
with values obtained in other $+U$ and hybrid-functional DFT
computations~\cite{rpapolaron,deskins}.


To elucidate the origin of the $A_{1g}$ mode stiffening, 
we calculated vibrational eigenmodes and frequencies 
using the frozen phonon method. Before performing
frozen-phonon calculations in the periodic cell containing an extra-electron,
(to which we refer as the \emph{`E'} simulation) and an electron-hole excitation
(to which we refer as the \emph{`E-H'} simulation),
we relax the geometry using a conjugate
gradient algorithm, starting from the last point of the corresponding
thermalized non-adiabatic trajectories described below. The results for the $A_{1g}$
and $E_{g}$ modes are presented in Table \ref{tab:freq}. 

\begin{table}
\begin{tabular}{|c|c|c|c|c|}
\hline 
\multirow{2}{*}{System} & \multicolumn{2}{c|}{$A_{1g}$ frequency, cm\textsuperscript{-1}} & \multicolumn{2}{c|}{$E_{g}$ frequency, cm\textsuperscript{-1}}\tabularnewline
\cline{2-5} 
 & Theory & Experiment & Theory & Experiment\tabularnewline
\hline 
\emph{GS} & 603 & 607 & 444 & 449\tabularnewline
\hline 
\emph{E} & 611 & - & 444 & -\tabularnewline
\hline 
\emph{E-H} & 616 & 617 & 446 & 440\tabularnewline
\hline 
\end{tabular}

\protect\caption{Frequencies of $A_{1g}$ and $E_{g}$ phonons obtained by theory 
and experiment. \emph{GS} refers to the ground state of titania, while
\emph{E }and \emph{E-H }to the extra-electron and photo-excited electron-hole systems.
\label{tab:freq} }
\end{table}

\begin{figure*}
\includegraphics[width=8.6cm]{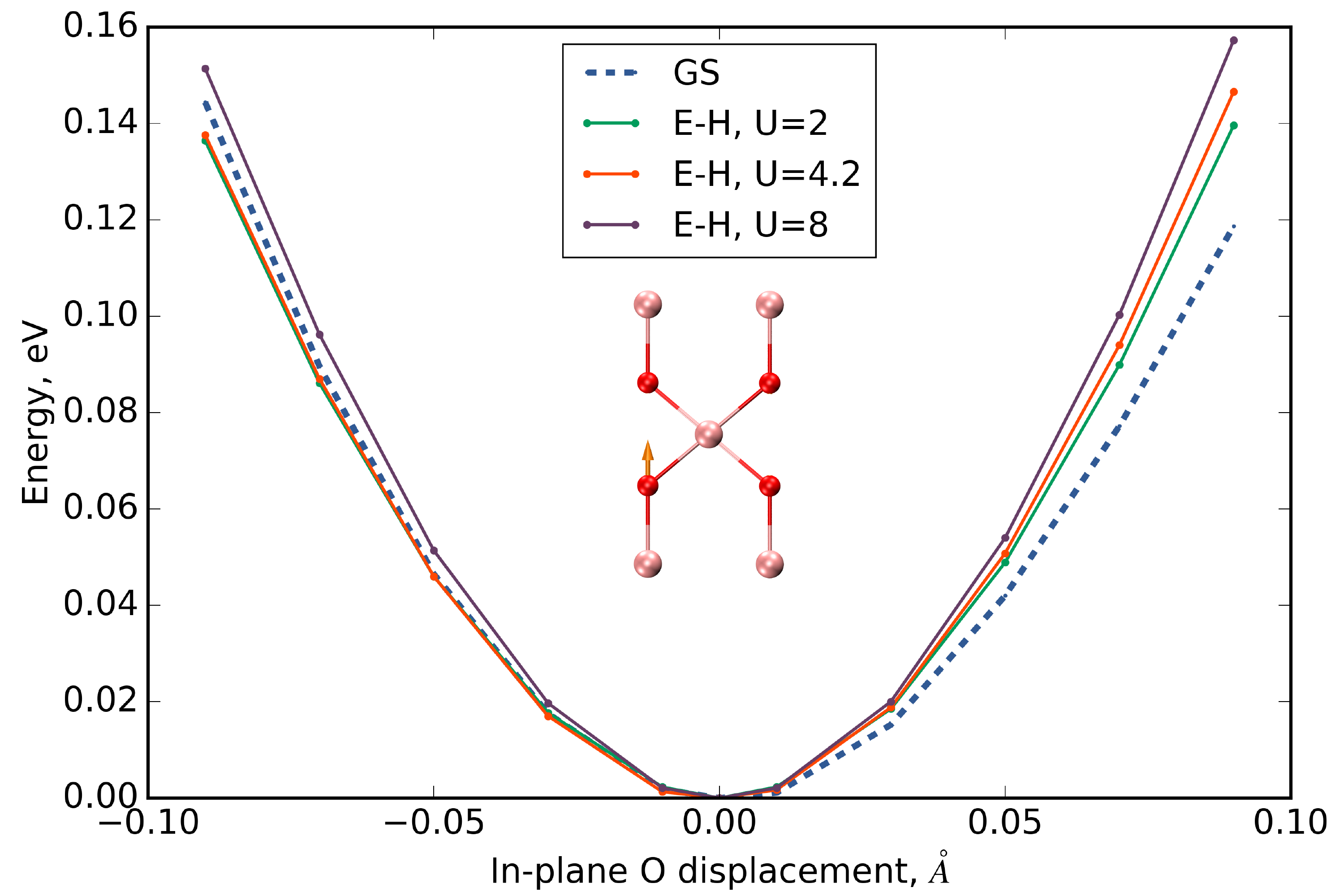}
\protect\caption{ 
Total electronic energy as a function of displacement of the O
atom along the component of the $A_{1g}$ eigenvector. Green, orange and purple
graphs corresponds to the exciton-polaron obtained for different values of $U$
and dashed blue line corresponds to the ground state at the ground state
geometry (positive displacement indicates diminished O-Ti$^*$-O angle).  
\label{fig:ushift}}
\end{figure*}

The agreement between experiment and theory is surprisingly good. 
Note that for the $E_{g}$ phonon the experimental peak is expected to red-shift
with increasing temperature~\cite{tio2_raman}, while in the simulation the
effect of heating is not included. Given the limitations of the experimental
measurements and of the simulations it could be argued that this surprising
level of agreement is due to fortuitous cancellation of errors. Because of the
agreement in the \emph{trend}, that is, the stiffening of the $A_{1g}$ mode
only, as well as systematic stiffening of this mode with increasing value of
$U$ (see Fig. \ref{fig:ushift} and text below),  we suggest that the simulation
captures, at least qualitatively, the mechanism of the stiffening, prompting us
to analyze further the vibrational and electronic structure of the
exciton-polaron system. 

 In Fig. \ref{fig:locA1g}c we show 
the atomic displacement corresponding to the $A_{1g}$ phonon in the \emph{E-H} simulation
(the phonon mode in the \emph{E }simulation
is nearly identical, see Fig. S2). 
Due to breaking
of the crystal symmetry (space group $P\frac{4_{2}}{m}\frac{2_{1}}{n}\frac{2}{m}$) 
by the lattice distortion associated with the polaron, the 616 cm\textsuperscript{-1}
phonon in the \emph{E-H }simulation (and the 611 cm\textsuperscript{-1}\emph{
} in the \emph{E} simulation) is completely localized around one of the Ti atoms, which we will denote as Ti\textsuperscript{{*}}. This new phonon is a symmetric oxygen breathing mode which belongs
to the $A_{1g}$ representation of the Ti$^*$ point
group symmetry $\frac{2}{m}\mbox{\ensuremath{\frac{2}{m}\frac{2}{m}}}$
($D_{2h}$). We denote this localized oxygen breathing mode as $A_{1g}^{*}$
for clarity. One of two degenerate $E_{g}$ modes (the one involving
apical O atoms bonded to Ti$^*$ as shown
in Fig. \ref{fig:locA1g}a  preserves its character upon formation
of the polaron, while the other (at 444 cm\textsuperscript{-1}) wraps
around the polaron and does not involve involve O atoms bonded to
Ti$^*$ (Fig. S3). We provide an explanation for this
effect below. The frequency of both $E_{g}$ modes is unchanged.

\begin{figure*}
\includegraphics[width=15cm]{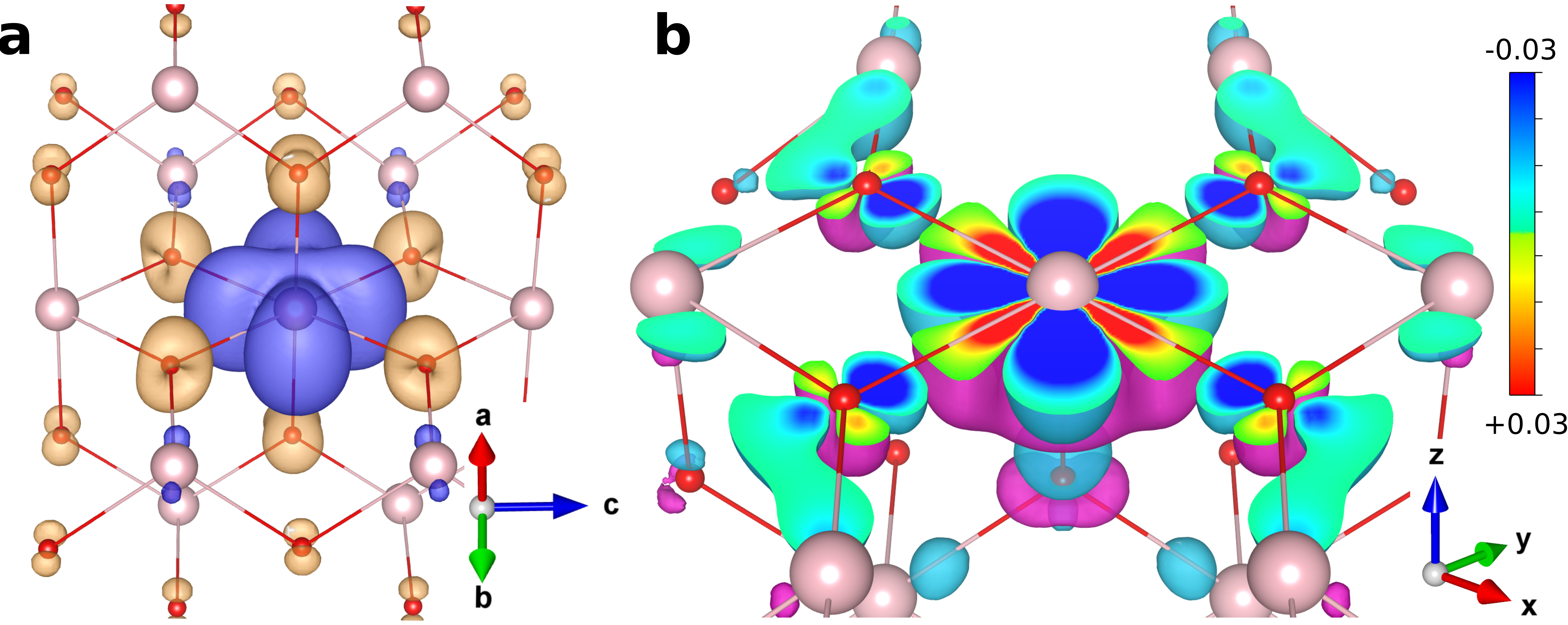}

\protect
\caption{Electronic structure of the exciton-polaron. (\textbf{a}) The
magnetization density: blue isosurface represents spin-up (electron) and orange
spin-down (hole) magnetization densities.  (\textbf{b}) Charge-density
difference between the electronic excited and ground states at the same atomic
geometry: magenta isosurface represents charge difference (excess of ``hole'')
and cyan negative (excess of ``electron'') charge difference .  We also show
the gradient of the density difference on the ($1\bar{1}0$) cut-through plane,
with blue-shades depicting excess electron density and red-shades depicting
excess hole density (see Fig. S4 for a similar picture on the (110) cut-through
plane); the cut-through planes are labeled according to crystallographic
directions.  \label{fig:el-polaron}}

\end{figure*}


In Fig. \ref{fig:el-polaron}a we show
the magnetization density associated with the excited electron (spin-up)
and the hole (spin-down). The excited electron is primarily localized
on the \emph{d\textsubscript{xy}} orbital of the Ti$^*$
atom, one of the three non-bonding $t_{2g}$ orbitals~
\footnote{Here for simplicity we use usual notation applied to octahedral environment. Because in rutile the octahedron is distorted the actual orbital structure and bonding is more complex\cite{tio2bonding,tio2bonding_xrd}}.
\nocite{tio2bonding, tio2bonding_xrd} 
It is partially screened by the hole which is weakly localized on
the next-neighbor O atoms. Most of the hole occupies delocalized \emph{p}-type
lone-pair orbitals of the O atoms in the system. The presence of the electron
in anti-bonding states and the hole in bonding states of the system,
as well as the mobility of both carriers, should result in an overall
softening of the potential. This simple picture ignores the rearrangement
of the total electron density caused by the localized electron.
In Fig. \ref{fig:el-polaron}b we present the electron density {\em difference}
between the exciton-polaron excited state and the ground state at the
same geometry. This plot shows that the localized electron causes
further polarization of the Ti$^*$-O bonds with
the electron on the 
$e_g$-like orbital 
displaced towards the O atoms.  This increase of valence electron density on O atoms compensates for
the presence of the hole; indeed, the net Hirshfeld atomic charge on the
oxygen atom is the same as in the ground state at the relaxed geometry.

We consider next the displacement of O atoms bonded to Ti$^*$ in the $\pm x,\mp y$
direction on the ($1\bar{1}0$) plane, see Fig. \ref{fig:el-polaron}b.
This corresponds to a scissor-like motion of four in-plane O atoms in the direction
of the $A_{1g}$ mode in Fig.\ref{fig:locA1g}a-b. 
The inward motion ($-x,+y$
direction for the O atoms in front and $+x,-y$ for the atoms in back) 
in the excited system should experience a harder potential
than in the ground state due to Coulomb repulsion between valence
electrons surrounding the O atoms and the excited electron in the
lobes of the \emph{d\textsubscript{xy}} orbital of the Ti$^*$
atom. Alternatively, this can be viewed as hardening due to the action of
the Hubbard \emph{U}with increasing overlap between the bonding \emph{p}-states
of O atoms and the \emph{d\textsubscript{xy}} orbital of Ti$^*$
and the increasing population of the latter state. In rutile titania the oxygen
octahedron is distorted and \emph{d}-orbitals belonging to the $T_{2g}$
(reducible) representation are not degenerate; thus the potential can not be
softened by moving the excited electron to another $t_{2g}$ orbital. Outward
motion of the O atoms ($+x,-y$ direction for the O atoms in front) is on the
contrary softened by the interaction with the excited electron as the atoms
move away from it and because of additional Ti-O bond rearrangement. This
change in potential leads to almost complete suppression of the scissor-like
motion of the four O atoms in the $xy$-plane for the $A_{1g}^{*}$ phonon, see
Fig. \ref{fig:locA1g}b. The effect of Coulomb repulsion is confirmed on Fig.
\ref{fig:ushift}, where the potential for the O atom is mapped for different
values of $U$ and displays systematic stiffening with increasing $U$
(graphs for $U=1$ and $U=6$ on Fig.\ref{fig:ushift}, which conform to the same
trend, were left out for clarity).  A similar description can be applied to the
apical O atoms vibrating along the Ti$^*$-O bond (along the
$z$-axis in Fig.  \ref{fig:el-polaron}), with the reduced effect of Coulomb
repulsion due to the absence of lobes of the occupied
\emph{d\textsubscript{xy}}-orbital in the $z$-direction (see Fig. S4 and Fig.
S5b). The overall effect is stiffening of the potential.  Similarly,
displacement in the direction of one of the \emph{E\textsubscript{g}} modes,
Fig.  \ref{fig:locA1g}a, does not change significantly the overlap of O
\emph{p}-states and the occupied \emph{d\textsubscript{xy}}-state of Ti$^*$ and
thus has a smaller effect on this phonon mode.

\begin{figure*}
\includegraphics[width=15cm]{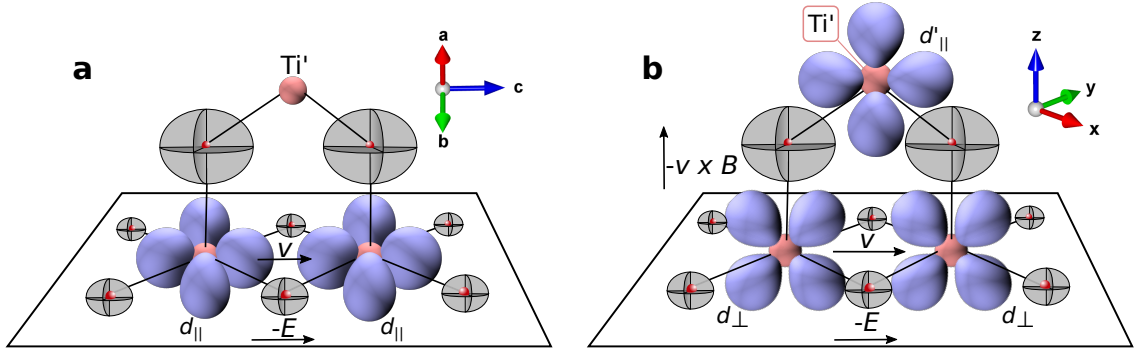}
\protect\caption{Scheme of polaron motion in Hall experiment.
(\textbf{a}) The polaron is hopping from a Ti$^*$ atom on the left to a Ti$^*$ atom  on
the right in the xy-plane, with the electron occupying their overlapping in-plane
lowest-energy $d_{||}$ ($d_{xy}$) orbitals. The electric field is in
the [$00\bar{1}$] direction. Thermal ellipsoids around the oxygens depict their
relative vibrational amplitude. (\textbf{b}) Same as (\textbf{a}) with
addition of magnetic field; the Lorentz force acts normal to the plane. The
Hall current is directed toward  orbital $d'_{||}$ on a Ti$'$ atom through the
intermediate orbitals $d_\perp$ on Ti$^*$ atoms. The vibrational displacement of apical O-atoms shown by ellipsoids elevates the energies of either $d'_{||}$ or $d_{\perp}$ leading to reduced Hall current.  \label{fig:hall}}

\end{figure*}

The asymmetry of the potential in relation to O motion and Ti$^*$ $t_{2g}$
orbitals may have interesting consequences for polaron transport in the
presence of a magnetic field. In Fig.  \ref{fig:hall}a we show the polaron
moving in the c-direction.  The electron is hopping from the lowest-energy
in-plane $d_{||}$ ($d_{xy}$) orbital of Ti$^*$ on the left to the same orbital
of the next Ti$^*$ on the right. The higher the population of phonon modes that
involve in-plane O-atoms the higher the rate of nonadiabatic hopping. High
population of the phonon modes causes larger displacement for the apical
O-atoms than for in-plane O-atoms, because the potential for the apical O-atoms
is softer.  These large displacements decrease the rate of hopping in
c$_{\perp}$ direction, because the large displacement  of negatively charged
apical O-atoms leads to increased Coulomb repulsion either by destination
orbital $d'_{||}$ ($d_{x'y'}$-orbital in the symmetry-transformed
$x,y,z$-coordinates)  or by relevant intermediate orbitals belonging to Ti$^*$
octahedra, $d_{\perp}$ (\emph{e.g.} $d_{xz}+d_{yz}$), through which such
hopping must proceed. This effect opposes the Lorentz force that drives the
Hall current, see Fig. \ref{fig:hall}b.  The same argument applies to the
motion of the polaron in the c$_\perp$-direction, with the Lorentz force acting
in the c-direction.  The higher the temperature, the higher the population of
the polaronic phonon modes, and therefore the higher the drift velocity, the
stronger this effect should be. This explains the observed decrease of the Hall mobility
with increasing temperature, while the drift mobility (hopping rate) increases. 

\section{Formation of the polaron and exciton-polaron}

\begin{figure}
\includegraphics[width=8.6cm]{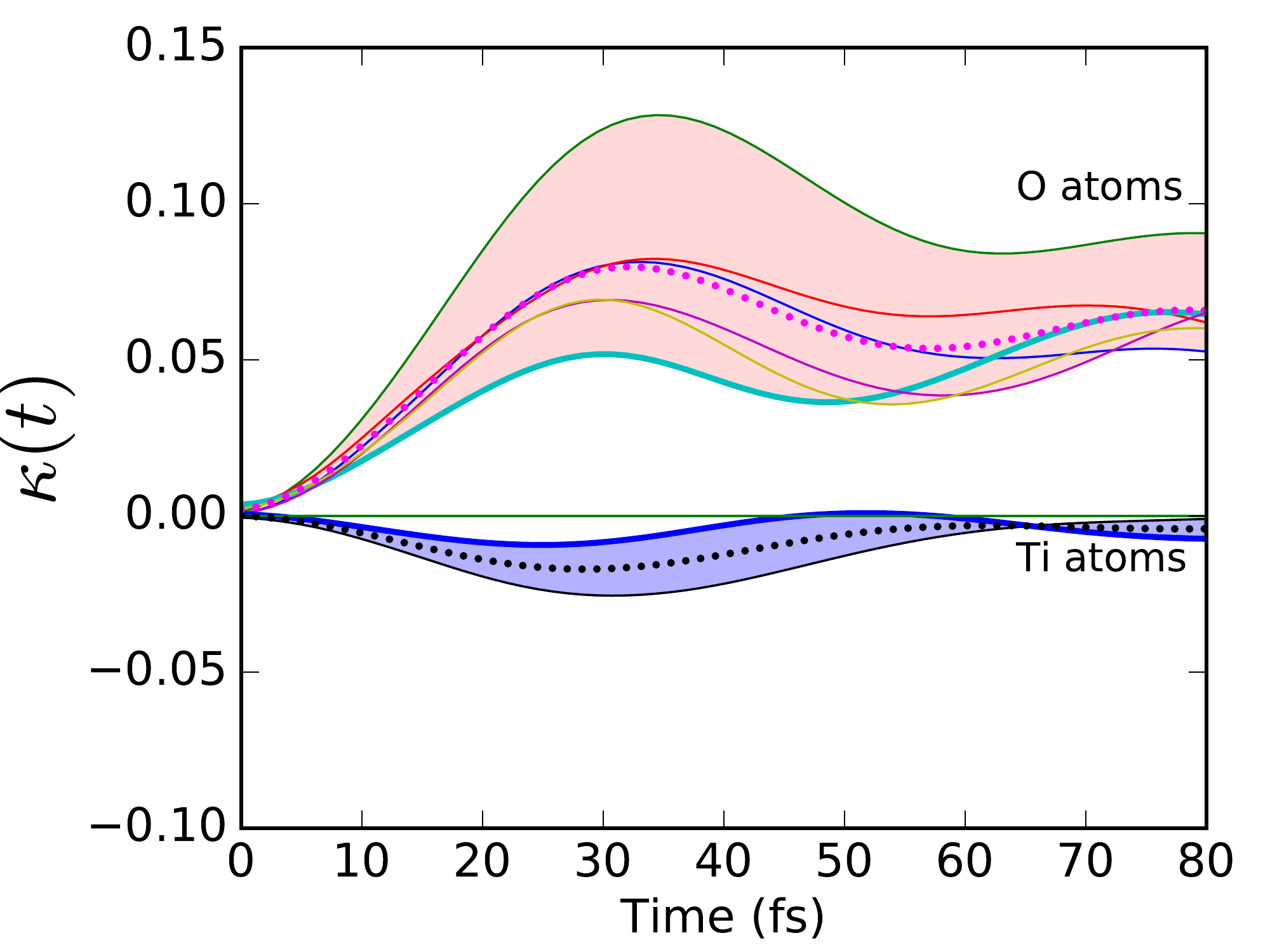}

\protect\caption{Lattice distortion associated with formation of the 
polaron for the photo-excited electron-hole (\emph{E-H}) non-adiabatic trajectory measured in terms of $\kappa$ Eq. (\ref{eq:cumaveragediff}),
the cumulative average differences from the ground-state-trajectory values 
of the distances between the Ti$^*$ atom and its nearest-neighbor
O and Ti atoms. The dotted lines correspond
to the average value of $\kappa$ for each atomic species.
\label{fig:hirshgelly}
(the corresponding plot for the $E$-trajectory is presented in Fig. S6).}
\end{figure}

We examined how the polaron is formed in two different simulations. In the
first case we introduced a single extra electron into the thermalized
($T=$300~K) system (\emph{`E'} simulation). In the second case we introduced an
electron-hole (e-h, \emph{`E-H'} simulation) pair by finding the lowest excited
state with the $\Delta$SCF method.  The \emph{E-H }trajectory is closer to the
experimental situation, where e-h pairs are generated by UV light. This choice
implicitly assumes that at experimental time scales the majority of the excited
electrons relax to the bottom of the conduction band and the holes to the top
of the valence band. Prior to introducing the e-h pair we thermalized the
system at $T=300$ K using standard Born-Oppenheimer molecular dynamics (BOMD).
To make the comparison between the two simulations more informative, we start
both at the same point of the BOMD trajectory, that is, with the same ionic
coordinates and velocities. These simulations were performed with Ehrenfest
dynamics within RT-TDDFT. As control samples, we used two BOMD trajectories
branching from the same original configuration, of which the first corresponded
to a neutral ground-state trajectory (\emph{GS} trajectory) and the second to
an adiabatic trajectory with an added electron (`\emph{ad-E'}). In both
non-adiabatic simulations we observed the formation of a polaron (Fig.
\ref{fig:hirshgelly}, Fig. S6 and movie S1). In the \emph{ad-E} simulation,
even though we are using the same small time-step as in the non-adiabatic
simulations ($\approx$24 as), we observed large deviations from
energy-conservation which lead to divergence of the trajectory in less than 5
fs.  This was caused by violation of the conditions of the Born-Oppenheimer
approximation: we assume that before the full polaron is formed, there are
states with the electron localized on different Ti atoms that are close in
energy, while interaction between them is small in comparison to the potential
barrier separating them.  This confirms that, at least in this case, the
formation of the polaron is indeed a non-adiabatic process.

A polaron consists of a localized charge trapped in a self-induced lattice deformation.
To quantify the deformation, for each nearest-neighbor $J$ of Ti$^*$ we compute
the quantity $\kappa_{J}$, defined as 

\begin{equation}
\kappa_{J}(t_{n})=\sum_{k=1}^{n}\left[d_{J}\left(t_{k}\right)-d_{J}^{GS}(t_{k})\right]/n,\label{eq:cumaveragediff}
\end{equation}

where $t_{n}$ is the time at time step $n$, $d_{J}$ is the distance between
Ti$^*$ and atom $J$ in the non-adiabatic trajectories
(\emph{E} and \emph{E-H}) and $d_{J}^{GS}$ is the same distance in the
\emph{GS} BOMD trajectory. We present in Fig. \ref{fig:hirshgelly} the values
of $\kappa_{J}(t)$ for different species in the neighborhood of
Ti$^*$. In both the \emph{E} and \emph{E-H} simulations, the
deformation converges at roughly 50-60 fs, which matches nicely the 55 fs
period of the $A_{1g}$ mode obtained in Raman experiments here and in Ref.
\citenum{tio2_raman}. This result is expected from small polaron
theory~\cite{holstein1959p2,devreese_review} and demonstrates that the
Ehrenfest approximation captures well the dynamics of small polaron formation.
As expected, negatively-charged O atoms are repelled from the Ti$^*$ atom in
both trajectories and the O-Ti$^*$ bond length increases substantially (by
$\sim0.07$~\AA).

\section{Conclusion}



Most theoretical polaron models employ a single-particle picture for the
electrons and linear electron-phonon interaction terms  in order to understand
carrier behavior and to interpret experimental results.  From such models,
softening of the polaron-renormalized phonon frequency $\tilde{\omega}$ is
expected, for any strength of electron-phonon interaction
\cite{alexandrov_strong,alexandrov_intermediate,lang_firsov,rashba1957,migdal,emin}.
In contrast to this, we observe stiffening of the $A_{1g}$ mode, both in the
experimental Raman spectra and in our DFT simulations.  Our analysis of the
simulations reveals strong anharmonic effects: upon breaking of the crystal
symmetry by the polaron, the original crystalline $A_{1g}$ mode is transformed
into a new oxygen breathing mode which is localized on the polaron.  The
formation of the polaron is a strongly nonlinear process that involves
significant change of the valence electron density around the localized
electron. In the simulation, the localization and rearrangement is caused by
the Coulomb interaction correction introduced by the Hubbard $U$ term in the
DFT$+U$ method. These effects cause overall stiffening and localization of the
Raman-active $A_{1g}$ phonon mode.  The resulting potential is asymmetric with
regard to motion of in-plane and apical oxygens, as well as to $t_{2g}$ and
other orbitals in the distorted octahedron. This observation led us to propose
a qualitative explanation of the anomalous temperature dependence of the Hall
mobility, which decreases with temperature, while the drift mobility increases
(above an activation temperature) as expected from non-adiabatic small-polaron
theory. The asymmetry of the potential leads, upon excitation of the phonon
modes required for polaron motion in one direction, to higher-amplitude motion
of the out-of-plane oxygens, which, through Coulomb repulsion, increase the
energy of the orbitals involved in hopping in the direction perpendicular to
the drift current, thus reducing the Hall current.  As a first step in
quantitative modeling of these problems we applied Ehrenfest dynamics with
RT-TDDFT to study the formation of the polaron in TiO\textsubscript{2} and
demonstrated the usefulness and applicability of this quantum-classical method
in describing the dynamics and the timescale of this process. These findings
may be relevant to other $d$- and $f$- materials which display small-polaron
properties.

\section*{Acknowledgments}
This work was supported by an NSF grant 
EFRI 2-DARE: Quantum Optoelectronics, Magnetolectronics and Plasmonics in
2-Dimensional Materials Heterostructures, Award 1542807.
Computational resources were provided by XSEDE~\cite{xsede} (Grant No.
 TG-DMR120073), which is supported by National Science Foundation Grant No. ACI-1053575, NERSC (ERCAP
request number 88881) and the Odyssey cluster, supported by
the FAS Research Computing Group at Harvard University.
\bibliographystyle{naturemag}

%


\end{document}